# A Three-Tiered Hierarchical Computational Framework Bridging Molecular Systems and Junction-Level Charge Transport

*Xuan Ji\*, Qiang Qi, Yueqi Chen, Chen Zhou, Xi Yu\**

Key Laboratory of Organic Integrated Circuit, Ministry of Education & Tianjin Key Laboratory of Molecular Optoelectronic Sciences, Department of Chemistry, School of Science, Tianjin University, Tianjin 300072, China

**ABSTRACT** The Non-Equilibrium Green's Function (NEGF) method combined with ab initio calculations has been widely used to study charge transport in molecular junctions. However, the significant computational demands of high-resolution calculations for all device components pose challenges in simulating junctions with complex molecular structures and understanding the functionality of molecular devices. In this study, we developed a series of approximation methods capable of effectively handling the molecular Hamiltonian, electrode self-energy, and their interfacial coupling at different levels of approximation. These methods, as three-tiered hierarchical levels, enable efficient charge transport computations ranging from individual molecules to complete junction systems, achieving an optimal balance between computational cost and accuracy, and are able to addresses specific research objectives by isolating and analyzing the dominant factors governing charge transport. Integrated into a Question-Driven Hierarchical Computation (QDHC) framework, we show this three-tiered framework significantly enhances the efficiency of analyzing charge transport mechanisms, as validated through a series of benchmark studies on diverse molecular junction systems, demonstrating its capability to accurately and efficiently elucidate charge transport mechanisms in complex molecular devices.

## 1. Introduction

Significant progress has been made in the research of molecular junctions and single-molecule devices in recent years,[1–6] thanks to progresses in characterization techniques, such as STM-BJ, and methods for fabricating robust single-molecule junctions like the development of graphene nanogap electrodes.[7] Mechanisms of charge transport across molecules and origins of molecular device functionalities, arising from the intrinsic properties of molecules, molecular-electrode interface characteristics, and electrodes, are continuously unveiled.[1,4,7–14]

Theoretical investigations have played a pivotal role in clarifying transport mechanisms within molecular junctions.[15] Quantum transport theory based on non-equilibrium Green's function formalism (NEGF) combined with density-functional theory (DFT) was originated with the tight-binding-based investigations by V. Mujica and M.A. Ratner in the mid-1990s[16], then extended to a non-self-consistent equilibrium DFT scheme by C. K. Wang *et al.*[17], and eventually advanced into a fully self-consistent NEGF-DFT framework by H. Guo, Q. Xue, and K. Stokbro in early 2000s, respectively.[18–21] Over time, this methodology has matured as numerous computational tools such as QuantumATK[22], NanoDCAL[18], TranSIESTA[23,24], SMEAGOL[25], etc. Although the DFT+NEGF approach provides detailed insights into the intrinsic properties of junctions with ab initio full-atomic precision[18,26], these calculations are resource-intensive, particularly when modeling semi-infinite electrodes and molecule-electrode interfaces. The computational demands escalate further when examining more complex molecular systems for functional devices, leading to significant time and resource expenditures.

In contrast, theoretical models based on simplified physical principles—such as individual electron states with electrode interface interactions —offer a more intuitive picture of charge transport.[12,27–30] These models are computationally economical and elucidate the fundamental physical essence of complex junction systems.[30–33] However, they often sacrifice chemical detail and rely on empirical or non-ideal experimental data for parameter determination, rendering them phenomenological and less effective for studying structure-property relationships.[34]

Charge transport in molecular devices is a multiscale physical challenge, dominated by the interactions between finite molecules (acting as scattering centers) and infinite electrodes (leads). This necessitates a balanced computational strategy that employs appropriate approximations and computational precision to efficiently and accurately model molecules, electrodes, and their interfaces, tailored to specific research goals. This perspective is mirrored in the diverse theoretical methodologies employed by computational scientists, ranging from comprehensive DFT+NEGF computations to single-state models as described above. These methodologies vary in system size assumptions and self-energy precision, from simple molecular scattering regions to the "extended molecule" concept necessary for exploring molecule-electrode interactions. Computational packages like AITRANSS in FHI-aims use large finite clusters as surrogate electrodes connected to an extended molecule with the wide-band limit (WBL) approximation.[35] Such strategies effectively model molecule-electrode interactions with reduced computational complexity, using simplified coupling matrices to efficiently compute the self-energy for the extended molecule.

Despite the extensive individual studies that have provided valuable insights, the current computational study on molecular junctions lacks systematic explorations within a unified methodological framework. The work by Verzijl *et al.*[36] stands out as a particularly instructive contribution: they explored a multi-level simplification scheme based on the wide-band limit (WBL) approximation, which has subsequently been flexibly applied by computational researchers for practical investigations.[37–39] However, most existing works address specific elements in isolation, such as electrode configurations or narrow computational method like the WBL, without integrating these findings into a broader, cohesive approach.

This work proposes an integrated, hierarchical computation framework for charge transport in molecular junctions with muti-scale approximations on the theoretical level.

Driven by the objectives of the research, i.e., the dominant factors in charge transport, we developed a three-tiered methodological approach. This involves the strategic construction of the Hamiltonian for the central closed system and the self-energy for the electrodes and the molecule-electrode interface, as illustrated in Figure 1. The framework, This Question-Driven Hierarchical Computation (QDHC) framework with its three levels of computational methods, achieves an optimal balance between computational speed and precision, and enriches the theoretical tools for the analysis of charge transport across molecules and simulation of molecular junctions and for the advancement of novel micro/nano devices.

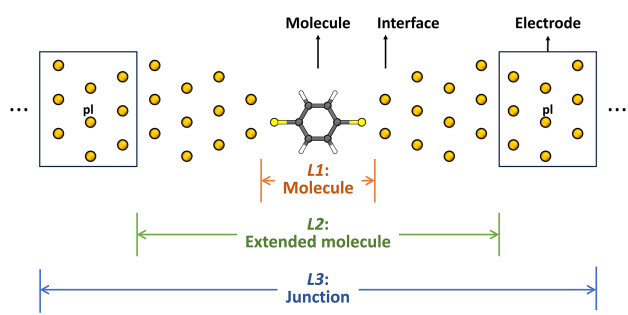

**Figure 1.** Schematic representation of the three-tiered methodological approach applied to the molecular junction within the hierarchical computation framework.

In subsequent sections, we will first present the theoretical foundation of this NEGF-based computation framework and then review the strategies for simplification and approximation commonly implemented in practical calculations. Attention will be paid on the scope of junction systems considered in the Hamiltonian construction, the theoretical level of Hamiltonian calculations, and methods for computing electrode self-energies. We then introduce our three-level computational framework, which progressively increases in complexity from molecular systems to complete junctions. Key advancements include self-consistent interfacial coupling calculations using size-optimized electrode clusters (Level 2), and efficient treatment of electrode electronic structures for accurate Fermi level alignment (Level 3), with bias transport modeled under a linear potential drop assumption. These hierarchical advancements distinguish our framework from previous implementations by providing a more balanced approach between computational efficiency and accuracy. Through systematic benchmark studies, we demonstrate how this this hierarchical design effectively captures the key characteristics of diverse molecular junction systems, providing targeted solutions based on specific research objectives.

## 2. NEGF formalism

The transport theory based on the non-equilibrium Green's function (NEGF) method follow a common framework based on the Landauer-Büttiker formalism,[29,40] by which the electric current $I$ through a junction is:

$$I = \frac{2e}{h} \int dE [f_L(E) - f_R(E)] T(E),  \quad (1)$$

where $f_L(E)$ and $f_R(E)$ are the Fermi-Dirac distributions for left and right electrodes, respectively. $T(E)$, the transmission probability, quantifying the likelihood of electrons at a specific energy $E$ traversing the junction from one electrode to the other, is given by:

$$T(E) = \text{Trace}[\Gamma_L(E) G^r(E) \Gamma_R(E) G^a(E)], \quad (2)$$

where $\Gamma_L$ and $\Gamma_R$ are the coupling matrices that describe how the junction interacts with the left and right electrode, respectively. $G^r$ and $G^a$ is the retarded and advanced Green's function of the transport region respectively, which capture the system's response to external disturbances by contacting with the electrode.

The Green's function is the solution to an equation of motion and physically reflects the particle propagation within the molecular junction system. The retarded Green's function is defined as:

$$G^r(E) = [(E + i\eta)I - H]^{-1}, \qquad (3)$$

where $+i\eta$ is positive infinitesimal imaginary term ensuring causality, meaning the response of a system only occurs in the future relative to a given impulse excitation at energy $E$. $H$ denotes the effective Hamiltonian of the transport system contacted with two semi-infinite leads, which reads,

$$H = H_0 + \Sigma_L + \Sigma_R \qquad (4)$$

where $H_0$ is the Hamiltonian of the isolated transport system, and the self-energy terms, $\Sigma_L$ and $\Sigma_R$, can be viewed as modifications of the Hamiltonian to involve the influence of semi-infinite electrodes. The self-energy is not Hermitian, and can be split into a Hermitian part according to

$$\Sigma_{L,R} = \Lambda_{L,R}(E) - \frac{i}{2}\Gamma_{L,R}(E) \qquad (5)$$

and the anti-Hermitian part

$$\Gamma_{L,R} = i\left[\Sigma_{L,R} - (\Sigma_{L,R})^\dagger\right] = -2\mathrm{Im}\Sigma_{L,R} \qquad (6)$$

which is also involved in Eq.2, quantify the broadening of molecular energy levels due to the interaction with the electrodes, reflecting how the electrodes' electronic states influence the molecule.

## 3. Computation Implementation of the NEGF

The practical computation of molecular junctions requires appropriate construction of the self-energies, $\Sigma$, for infinitely large electrodes, and the finite closed quantum system $H_0$, which in the context of molecular junctions, is often referred to as the scattering region (or the transport region). A key challenge in practical computational implementations of NEGF method lies in the difficulty in partitioning the molecular device into self-energy and transport region.

A straightforward/naive approach is to consider the molecule itself as $H_0$, based on the assumption that the molecule is typically the bottleneck of the electron flow. And all metal atoms that constitute the electrodes are entirely represented by the self-energies. While this model may suffer from oversimplification for the molecule-electrode interface, it provides a clear and concise description of molecular junctions, where the electric properties are mainly contributed by molecules. For such a simplified model, a constant self-energy was generally adopted for the lead (the electrodes) under the wide band limit (WBL) approximation.

$$\Sigma_{L,R} = \Lambda_{L,R} - \frac{i}{2}\Gamma_{L,R} \qquad (7)$$

Given the neglect of the level-shift by $\Lambda_{L,R}$, we obtain a self-energy of the form

$$\Sigma_{L,R} = -\frac{i}{2}\Gamma_{L,R} \tag{8}$$

The underlying assumption is the constant density of states of the electrode material around the Fermi energy, thus leading to an energy-independent coupling. It is well-suited for cases such as gold electrodes. Additionally, a weak interaction between the molecule and the electrodes is required to ensure that the electrode atoms do not alter the intrinsic properties of the molecule.

In practical molecular junctions, however, atoms within the nanogap of the electrodes often detach, forming pronounced tips that interact with the molecule, thus creating a junction.[41,42] The transport properties are determined not only by the electronic structure of the molecule but also by the presence of these atoms at the molecule-electrode interface.[43,44] These emergent electrode tips, which forming clusters, exhibit behavior distinctly different from that of idealized infinite electrodes, necessitating a different modeling approach. Additionally, the interactions between the molecule and electrode tips diverge markedly from those with infinite electrodes.[45] Therefore, a more rigorous strategy includes both the molecule and the electrode tips within the transport region, often referred to as the "extended molecule".[43,46–50] Moreover, regions of the electrodes near these tips, which do not exhibit significant protrusions, should be carefully examined due to their interaction with the extended molecule. As a result, these areas must also be incorporated into the modeling system to accurately represent the transport properties.[35,50–52]

This approach takes into account the influence of the semi-infinite electrodes, where electron transitions occur between discrete energy levels of the extended molecule and a continuum of energy states ($k$) therein, impacting the self-energy term.[53,54] The imaginary part of the self-energy can be expressed as:

$$\Gamma_{L/R}(E) = 2\pi \sum_k \left|V_{L/R}^k\right|^2 \delta(E - E_k) \quad (9)$$

This expression is intrinsically relevant to the electrode's density of states (DOS), $D(E) = \sum_k \delta(E - E_k)$, which is pertinent in situations where the WBL approximation does not hold.[55] Such scenarios include systems with non-gold metal electrodes, semi-metal electrodes like bismuth, and low-dimensional materials. The self-energy terms can be computed in practice to encodes the influence of the extended molecule's interaction with the infinite electrodes by the following formula:

$$\Sigma_{L/R}(E) = V_{L/R}^\dagger G_{\text{elec}}^{\text{surf}}(E) V_{L/R} \quad (10)$$

The surface Green's function (SGF) $g_{\text{elec}}^{\text{surf}}$ is practically used to model the effects of semi-infinite electrodes, effectively transforming these effects into boundary conditions that simulate local interactions within the transport region.

On the other hand, $H_0$ serves as the foundational Hamiltonian, defining the baseline properties such as energy levels and quantum states within the transport region. The computation of $H_0$ typically involves various quantum chemistry calculations, ranging from relatively simple approaches like the tight-binding or Hückel model—suitable for smaller systems or systems with straightforward electronic interactions with minimal computational

cost[56–58]—to more sophisticated techniques such as Density Functional Theory (DFT), which offers a practical balance between computational feasibility and quantitative accuracy.[59,15,60,61] For systems requiring a more rigorous treatment of electron-electron interactions, advanced many-body techniques, such as the GW approximation, can be employed. Each of these methodologies presents distinct trade-offs between computational cost and the accuracy of the resulting system description.

In the case of extended molecular systems, a broader and more comprehensive formulation of $H_0$ is used to self-consistently incorporate the effects of interface contacts on the molecular electronic states. This consideration is crucial when the coupling between the molecule and electrodes leads to significant modifications in the eigen states. However, as the size and complexity of the transport region increase, the associated computational burden also escalates, necessitating the development of more efficient computational strategies or the adoption of model simplifications to effectively manage these costs while retaining an acceptable level of accuracy.

## 4. The QDHC Framework

Our computational framework employs a hierarchical computational approach that balances the size of the transport region and the precision of self-energy treatments. This structure enables a rational, problem-driven methodology for studying charge transport in molecular devices, aiming for an optimal trade-off between computational cost and accuracy. The framework is organized into three levels—L1, L2, and L3—which correspond to different scopes within a molecular junction, as depicted in Figure 1. Each level progressively enhances precision and complexity, allowing the molecule, electrodes, and their interfaces to be treated with varying levels of detail, depending on their significance to the system.

## 4.1. L1-Molecule

Level 1 focuses on scenarios where the molecule is the dominant factor in charge transport within the molecular junction.[57] In this case, the intrinsic properties of the molecule are crucial, while the electrodes and their interactions with the molecule can be treated under WBL approximation. To accommodate molecules of varying complexity, L1 incorporates multi-precision computational methods, including empirical extended Hückel molecular orbital method, charge self-consistency processes[62], and the more universally applicable GFN-xTB method[63]. On the other hand, under the WBL approximation, the molecule-electrode interaction was modeled as constant coupling, expressed as:

$$\Sigma_{L,R} = -\frac{i}{2}\Gamma_{L,R} \qquad (11)$$

In this formulation, the $\Gamma$ and $\Sigma$ matrices correspond to the size of the Hamiltonian of the molecule, with non-zero elements only along the diagonals corresponding to the atomic orbitals at the two coupling sites, as depicted in Figure 2. This simplification allows the shape of the transmission functions to intuitively reflect the intrinsic electronic properties of the molecule[58,64], like quantum interference, Fano resonance, and molecular conformational effects, etc., thereby streamlining the analysis.

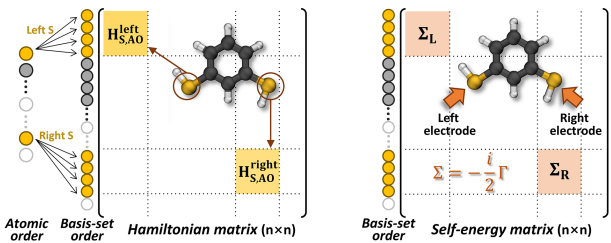

**Figure 2.** Schematic representation of the Level 1 (L1) computational approach, using the 1,3-BDT molecule as an example to illustrate the construction of the Hamiltonian matrix and the application of self-energy.

## 4.2. L2-Extended molecule

Level 2 builds upon the "extended molecule" configuration, which incorporates pronounced electrode tips, interacting with the molecule within the nanogap, into the transport region. These deliberately constructed interfaces account for the geometric characteristics of the tips and their connection patterns with the molecule, therefore effectively capture phenomena such as state hybridization and the dipole effects that stem from charge transfer between the electrodes and the molecule. Such refinements greatly enhance the realism of the molecular device models. Additionally, regions of the electrodes near these tips, must be carefully examined, as their interactions encapsulate the main influence of the infinitely large electrodes on the extended molecule.[52] These electrode regions are modeled as finite clusters and incorporated into the extended molecule for the total Hamiltonian and overlap matrices for the entire "**L cluster – extended molecule – R cluster**" configuration, as shown in Figure 3a. The expression is:

$$\boldsymbol{H}_{\text{sys}} = \boldsymbol{H}_{\text{L cluster}} \oplus \boldsymbol{H}_{\text{EM}} \oplus \boldsymbol{H}_{\text{R cluster}} = \begin{bmatrix} \boldsymbol{H}_{\text{L}} & \boldsymbol{V}_{\text{LM}} & \\ \boldsymbol{V}_{\text{ML}} & \boldsymbol{H}_{\text{EM}} & \boldsymbol{V}_{\text{MR}} \\ & \boldsymbol{V}_{\text{RM}} & \boldsymbol{H}_{\text{R}} \end{bmatrix} \quad (12)$$

Here, $\boldsymbol{H}_{\text{sys}}$ is the Hamiltonian matrix of the entire system for SCF calculation; $\boldsymbol{H}_{\text{L}}$ and $\boldsymbol{H}_{\text{R}}$ represent the Hamiltonians of the left and right electrode clusters, respectively; $\boldsymbol{H}_{\text{EM}}$ is the Hamiltonian of the extended molecule; the four nearest-neighbor off-diagonal matrices are the couplings between the electrode clusters and the extended molecule. A similar expression exists for the overlap matrix $\boldsymbol{S}_{\text{sys}}$. The direct interaction between the left and right electrode is negligible. By partitioning $\boldsymbol{H}_{\text{sys}}$ (also $\boldsymbol{S}_{\text{sys}}$) into subblocks corresponding to the dimensions of the electrode clusters and the extended molecule, we can self-consistently evaluated their coupling matrices. We draw upon the method proposed by Liu and Neaton[55], who demonstrated the effectiveness of such matrix decompositions capturing subtle electrode-molecule interactions. This approach yields a more accurate representation of self-energy terms beyond the WBL approximation:

$$\boldsymbol{\Sigma}_{\alpha=\text{L,R}}(E) = [E\boldsymbol{S}_{\text{M}\alpha} - \boldsymbol{V}_{\text{M}\alpha}]\boldsymbol{g}_\alpha[E\boldsymbol{S}_{\alpha\text{M}} - \boldsymbol{V}_{\alpha\text{M}}] \quad (13)$$

where $\boldsymbol{g}_\alpha$, the surface Green's function for the electrodes, is approximated as that of the bulk electrode. With a further constant local density of states (LDOS) assumption for the lead, justified by minimal variations in electronic states near the Fermi level, especially LDOS = $\rho(E) = 0.036 \; eV^{-1}$ for gold.[65] $\boldsymbol{g}_\alpha$ is therefore characterized by a constant imaginary component and represented as a diagonal matrix:

$$(\boldsymbol{g}_\alpha)_{ij} = -i\pi \text{LDOS}^{\text{const}} \delta_{ij} \quad (14)$$

Finally, the retarded Green's function for transmission is computed over the extended molecule region:

$$G_{\text{EM}}^{r}(E) = \left[(E + i\eta)S_{\text{EM}} - H_{\text{EM}} - \Sigma_{\alpha=\text{L,R}}(E)\right]^{-1} \quad (15)$$

This approach is illustrated in Figure 3a. Furthermore, when analyzing transport properties that necessitate Fermi energy, the HOMO of the electrode cluster can be utilized as a proxy, providing a reasonable approximation.[66]

Although similar molecular junction modeling strategies have been employed in previous studies, the primary challenge is determining the optimal size of the finite cluster (a small group of atoms used to approximate larger systems) that captures sufficient interaction with the extended molecule, while still minimizing computational cost. Previous research[35,67] have shown that interactions between the molecule and the electrode are generally localized, meaning they only affect the portion of lead atoms in direct contact with the molecule. Therefore, in principle, it is only necessary to calculate a finite cluster of lead atoms.

To identify the minimum cluster size that adequately represents the system, we investigated a 1,4-benzenedithiol (1,4-BDT) molecular junction connected to gold trimers, which were further attached to oversized clusters (each containing 108 gold atoms). This choice is based on following considerations:

1) Gold is the most widely used electrode materials in molecular junction study due to its superior electrical properties and chemical stability. This widespread use in both experimental and computational research has generated an extensive database of reference materials, facilitating comprehensive comparisons and result validation.

2) Gold exhibits relatively flat electronic band structures near the Fermi energy, making it particularly suitable for applying the wide-band limit (WBL) approximation in theoretical modeling.

3) The chemical bond between sulfur and gold is among the strongest in molecular junctions.[68] Therefore, the interaction range between thiols (sulfur-containing compounds) and gold electrodes is expected to be maximized, allowing us to establish an upper bound for the minimum cluster size.

4) 1,4-BDT is one of the most extensively studied systems in charge transport research, serving as a standard benchmark molecule for both theoretical and experimental investigations. Its well-documented electronic properties and abundant literature data make it an ideal model for validating computational methods and comparing experimental results across different studies.[35,45,69]

5) The trimer configuration represents the predominant structure for molecular anchoring to gold surfaces. In addition, we investigated the pyramid configuration as it is widely used in molecular junction studies and provides enhanced anchoring stability. This dual configuration validates the robustness of our approach, as the minimum cluster size derived from the trimer configuration remains applicable to the larger pyramid structure, demonstrating the broader applicability of our findings across different electrode configurations.

When examining the coupling matrix that describe the mathematical connections between different parts of the system (Figure 3b), and the self-energy terms that describe the influence of the environment on the system (Figure 3c), we indeed observed highly localized interactions between the electrode and the extended molecule. Significant non-zero matrix elements were observed exclusively at the interface between the electrode cluster and the trimer gold atoms of the extended molecule. These localized interactions suggest that strategic matrix truncation, through the elimination of negligible elements, can significantly reduce computational costs while maintaining accuracy. We implemented this truncation by setting matrix elements below a specified energy threshold to zero, thereby preserving only the significant interactions with metal atoms at contact points. In the case of 1,4-BDT, we identified the maximum coupling element of 0.62 eV at the sulfur atom and subsequently established a cutoff threshold of 0.65 eV. The transmission spectrum computed using this truncated matrix (Figure 3e) showed excellent agreement with results from the full matrix calculation, validating that our reduced coupling region is sufficient for accurate transport calculations.

We thus identified a 25-atom cluster from the oversized electrode based on the modified coupling matrices with non-zero values for the trimer gold contacts. Further analysis presented in the Supplementary Information (Figure S1, S2) demonstrates that the localized interaction pattern remains consistent across various chemical anchoring groups, validating the broad applicability of our energy truncation approach.

A slight deviation was observed for the thiomethyl-terminated (-SMe) molecule, which exhibited enhanced coupling effects due to their reduced distance from the electrode surface. To address this case, we employed a pyramid structure with an apex atom serving as the interface. The previously established 0.65 eV energy cutoff proved suitable for this configuration, as detailed in the Supplementary Information (Figure S3, S4). Our results thus indicate that both the trimer and pyramid structures, when implemented with a 25-atom gold cluster, achieve an optimal balance between computational efficiency and numerical accuracy.

We further investigated the effective cluster size for the adatom structure, another prevalent interface configuration commonly observed in molecular junctions with amine- or pyridine-anchoring group. Using the same approach, we determined an optimal cluster size of 15 atoms, as shown in Figure 3g and detailed in the Supplementary Information (Figure S5, S6).

In summary, at Level 2, we expanded the transport region to include the "extended molecule" and improved the self-energy treatment by accounting for localized interactions between the extended molecule and electrode clusters. By implementing a coupling energy-based cutoff criterion, we established optimized electrode cluster dimensions and interface configurations that effectively balance computational accuracy with numerical efficiency. Our investigation provides a comprehensive framework for selecting optimal interface configurations and cluster dimensions, as detailed in Table 1. This reference table offers specific computational recommendations for diverse molecular systems, especially for prevalent anchoring groups. For example, we recommend a 25-atom trimer cluster for chemisorbed thiolate, amine, and pyridine groups, and a 25-atom pyramid cluster for these groups as well as thiomethyl. For systems anchored by amine and pyridine anchoring groups, a 15-atom adatom cluster serves as an efficient alternative.

**Table 1.** Summary of the Electrode Cluster type and recommend anchoring groups

| Interface type | Cluster size | Anchoring groups |
| --- | --- | --- |
| Trimer | 25 | Chemisorbed thiolate, amine, pyridine |
| Pyramid | 25 | Thiomethyl, chemisorbed thiolate, amine, pyridine, |
| Adatom | 15 | Amine, pyridine |

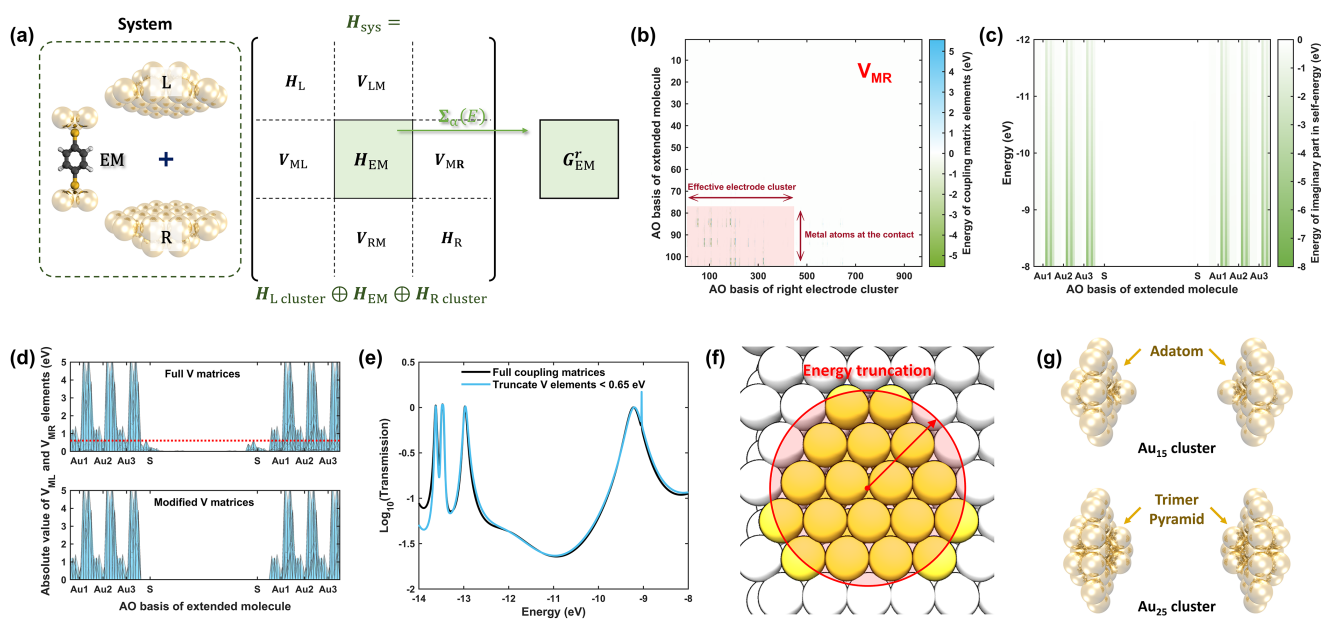

**Figure 3.** (a) Schematic representation of the Level 2 (L2) approach, featuring the extended molecule connected to electrode clusters. The Hamiltonian is computed for the entire system, partitioned into subregions, which then yields coupling matrices for self-energy calculation. These, along with the extended molecule's Hamiltonian, enable construction of the Green's function for transport calculations as detailed in the methodology section. (b) Coupling matrix $V_{MR}$ for the 1,4-BDT extended molecule with oversized electrode clusters, highlighting non-zero elements localized at the contact points between the metal atoms in the extended molecule and the adjacent regions of the electrode cluster. (c) Energy-dependent distribution of the imaginary components of self-energy matrices' diagonal elements, demonstrating spatial localization at metal atom sites. (d) Coupling matrices and (e) corresponding transmission spectra before and after implementation of the energy cutoff. (f) Optimized cluster region determined through application of the 0.65 eV energy cutoff threshold. (g) Recommended cluster sizes for common contact configurations based on our energy cutoff criterion.

### 4.3. L3-Full Junction

#### 4.3.1. Equilibrium Situation

Level 3 extends the L2 approach by incorporating a more sophisticated treatment of the electrodes and their coupling to the transport region, thereby providing a more rigorous description of the full molecular junction. As illustrated in Figure 4a, this is realized via a two-step treatment applied separately to the electrode and junction region. First, periodic boundary conditions (PBC) were applied in the calculation of bulk-phase properties to approximate infinite electrodes, enabling an accurate representation of their electronic band structure. Second, the interaction between the electrodes and the extended molecule was refined through a self-consistent computation of the junction system, which consists of the extended molecule and one principal layer from each electrode[36,70,71]. Particularly noteworthy is the work of Verzijl *et al.* [36], which explored a similar junction treatment but utilized the WBL self-energy at the interfaces, encompassing a less detailed description of electrode characteristics.

For the electrode modeling, we employed a unit cell containing two principal layers of Au(111), which is twice the content of the solid box shown in Figure 4a. Multiple k-points were sampled within the Brillouin zone along the transport direction, while a single k-point was used in the transverse directions, balancing between computational efficiency and numerical accuracy. The total and projected DOS profiles of bulk Au, provided in the Figure S8, further validate the accuracy of this setting. The resulting k-dependent Hamiltonian matrices were subsequently transformed to the real-space tight-binding representation via inverse Fourier transform and subsequently partitioned into sub-blocks describing each principal layer and their interactions:

$$\boldsymbol{H}_{\text{bulk}} = \begin{bmatrix} \boldsymbol{H}_{00} & \boldsymbol{H}_{10} \\ \boldsymbol{H}_{01} & \boldsymbol{H}_{11} \end{bmatrix} \qquad (16)$$

where $H_{00}$ and $H_{11}$ are identical in structure but differ in notation. The surface Green's function (SGF) at specific energy points was computed iteratively using the Lopez-Sancho method[72,73]:

$$H_{00} \rightarrow H_{00}^{\text{surf}} \tag{17}$$

$$g_{00}^{\text{surf}}(E) = \left[(E + i\eta)S_{00} - H_{00}^{\text{surf}}\right]^{-1} \tag{18}$$

And the density of states (DOS) can be obtained by,

$$\text{DOS}(E) = -\frac{1}{\pi}\text{Im}\left[\text{Trace}\left(g_{00}^{\text{surf}}(E)\right)\right] \tag{19}$$

While the projected DOS for each valence shell can be found in Figure S9, Figure 4d showed that the total DOS exhibits a distribution like that of bulk-phase Au, with a relatively flat profile near the Fermi energy, supporting the validity of the wide-band limit (WBL) approximation. However, at energy regions beyond the vicinity of the Fermi level, significant variations in the band structure of gold, primarily due to the 5d orbitals, can substantially influence the molecule-electrode interface interactions and, consequently, the transport properties. These variations necessitate an energy-dependent treatment of the electrode properties, superior to the LDOS approximation used in Level 2, to accurately capture these effects.

Considering that the bulk-phase electronic structure of the electrodes is stable and minimally affected by the molecular junction, it is unnecessary to recalculate the electrode SGF for each junction computation. Therefore, we pre-calculated the SGF at sparse energy points $\left(E_q, \boldsymbol{g}_{00}^{\text{surf}}(E_q), q = 1,2,\ldots,n.\right)$ and SGFs for all required energy points can then be generated through linear interpolation:

$$\boldsymbol{g}_{00}^{\text{surf}}(E_i) = \textbf{Interpl}\left(E_q, \boldsymbol{g}_{00}^{\text{surf}}(E_q), E_{\text{total}}\right), \ E_i \in E_{\text{total}} \tag{20}$$

As demonstrated in Figure 4d, the DOS obtained by interpolating from pre-calculated SGFs at 0.5 eV intervals primarily captures the sketchy shape of the DOS.

For the junction calculations, we proceeded with the configuration shown in Figure 4b, where the extended molecule is connected to a principal layer from each electrode. Like the L2 approach, the Hamiltonian (also overlap matrix) of the junction system is divided into three parts:

$$\boldsymbol{H}_{\text{junc}} = \boldsymbol{H}_{\text{L pl}} \oplus \boldsymbol{H}_{\text{EM}} \oplus \boldsymbol{H}_{\text{R pl}} = \begin{bmatrix} \boldsymbol{H}_{\text{L pl}} & \boldsymbol{V}_{\text{LM}} & \\ \boldsymbol{V}_{\text{ML}} & \boldsymbol{H}_{\text{EM}} & \boldsymbol{V}_{\text{MR}} \\ & \boldsymbol{V}_{\text{RM}} & \boldsymbol{H}_{\text{R pl}} \end{bmatrix} \tag{21}$$

This division represents the interactions between the extended molecule and the principal layers of the electrodes, and direct interactions between two principal layers at two sides of the junction are ignored. After linearly interpolating the pre-calculated SGF, the self-energy matrices at each energy point are calculated as follow:

$$\boldsymbol{\Sigma}_{\alpha=\text{L,R}}(E_i) = [E_i \boldsymbol{S}_{\text{M}\alpha} - \boldsymbol{V}_{\text{M}\alpha}] \boldsymbol{g}_{00,\alpha}^{\text{surf}}(E_i) [E_i \boldsymbol{S}_{\alpha\text{M}} - \boldsymbol{V}_{\alpha\text{M}}] \tag{22}$$

It offers several advantages in terms of self-energy accuracy over previous treatments that rely on the constant coupling approximation or assume a flat DOS for the electrodes. An accurate depiction of the electrode's electronic structure and its influence on transport is accounted for. The computational cost wherein is avoided by interpolating SGFs. This method is not limited to a specific electrode material; in principle, it can be applied to other electrodes with complex band structures, making our strategy more versatile and broadly applicable compared to the previous two levels.

Beyond L2, the additional bulk-phase calculations for the electrodes of L3 provide the bulk Fermi energy, $E_F^{\text{bulk}}$, enabling transmission features to be explored relative to the Fermi level. Since the principal layers in the junction region are structurally identical to those in the bulk electrode, their band features can be approximated by those of the bulk, only differing in chemical conditions. Figure 4c indicates a $\Delta\phi$ local potential mismatch between them, and alignment is achieved by adjusting the diagonal elements of the subblock Hamiltonian for the left/right principal layer ($\boldsymbol{H}_{\alpha\,\text{pl}}$) in the junction to match the bulk Hamiltonian ($\boldsymbol{H}_{\alpha\,\text{bulk}}$), thereby accounting for potential energy deviations[74]:

$$\Delta\phi = \frac{1}{2n_{\text{pl}}} \sum_{k\in\text{L,R}} \frac{(\boldsymbol{H}_{\alpha\,\text{bulk}})_{kk} - (\boldsymbol{H}_{\alpha\,\text{pl}})_{kk}}{(\boldsymbol{S}_\alpha)_{kk}} \tag{23}$$

By plotting the transmission spectrum relative to $E_F^{\text{bulk}} - \Delta\phi$, we can analyze transport properties tied to energy level alignment and identify the dominant conducting channels.

The computational procedure involved the following steps, aligning with the approach outlined above. First, bulk-phase electrode calculations specific to Level 3 were performed, but this step is not required when applying the method in practice if pre-calculated data is already available. Next, Hamiltonian matrices for the junction system were computed, and the system was divided into subregions of the extended molecules and the electrode clusters to extract coupling matrices between them. Combing these, energy-dependent self-energies were calculated by interpolating surface Green's functions (SGF), which had been precomputed at 0.5 eV intervals. Finally, the Hamiltonian of the extended molecule and self-energy matrices were combined to generate Green's function, then compute transmission spectra as a function of energy relative to the Fermi level.

It should be noted that due to the influence of the chosen computational method, may as well say the functional precision, and interface-induced renormalization effects, more accurate energy level alignment requires introducing an additional self-energy correction, as demonstrated by Neaton *et al.*[75–78] Although this procedure was not incorporated in the present work, it can be straightforwardly integrated into our computation framework to further refine the predicted energy levels and enhance the overall accuracy of the calculations.

### 4.3.2. Biased condition

With the Fermi level available, biased transport behaviors and current-voltage (I-V) characteristics can be further explored, like rectification, where current is evaluated by integrating the transmission spectrum within the bias voltage window centered at the Fermi energy. Under non-equilibrium scenarios, the external bias enters the calculations through not only shifting in the chemical potentials of the electrodes, but also inducing the polarization of electron density within the transport region, which reshapes the transmission functions. In principle, these effects are treated within the standard NEGF approach by imposing open boundary conditions corresponding to the electrochemical potentials in the electrodes, as well as by iteratively solving the Poisson equation for electrostatic potential and updating Green's function to achieve self-consistency of the electron density.

Instead, we adopt a surrogate approach by applying a uniform external electric field (EEF) along the transport direction within the transport region (the area marked as 'Junction' in Figure 4a) to generate the Hamiltonian perturbed linearly, and then constructing the Green's function in a single calculation. This approach assumes that the intrinsic electronic structure of the isolated transport region dominates the transmission properties. Although this approach does not enforce physically continuity of the electrostatic potential throughout the entire system, it is acceptable under small bias conditions where charge rearrangement effects are weak, and for higher bias, it provides an economic and acceptable approximation, as shown in previous studies[79].

When a voltage bias $V$ is applied to a molecular junction, it is generally expected that most of the electrostatic potential drop occurs at the molecule–electrode interfaces, where the resistance is largest.[80] Nevertheless, in the case of the molecule strongly coupled to the electrodes, the molecule backbone will also experience substantial potential differences, hence the potential drop is more evenly distributed across the extended molecule.[71,81] Considering this, we set the field strength as the bias voltage divided by the length of the extended molecule, resulting in a linear potential profile across the extended molecule, as depicted in Figure 4f. All other computational procedures remain consistent with those in the unbiased scenario. Our method therefore adequately captures the essential physics by directly modifying the electronic structure of the extended molecule.

We acknowledge that this is a quite phenomenological treatment that deviates from strict theoretical methods. Even so, for practical purposes and computational efficiency, assuming a non-self-consistent electron density and a linear potential drop remains reasonable and useful for quick qualitative analysis in low-bias transport cases.

In summary, compared to Level 2, the Level 3 approach simulates the characteristics of a full molecular junction more accurately through two key advancements. First, by performing bulk electrode calculations and iterating over surface states, Level 3 approach captures the energy-dependent features of the electrode's band structure, allowing for a more precise description of molecule-electrode interactions. The iteration processes is reduced by interpolating the pre-calculated SGFs, thus enhancing efficiency, and this can extend beyond traditional gold electrodes, making the framework applicable to various electrode materials. Second, by calculating the junction configuration where the extended molecule is connected to electrode principal layers, we can isolate the electronic properties of the bulk electrodes, allowing for the transmission spectrum to be aligned with the electrodes' Fermi energy. This approach enables more detailed investigations into nonequilibrium transport phenomena, such as the transport under bias (uniform EEF), not achievable using the cluster-based configuration employed in the previous level.

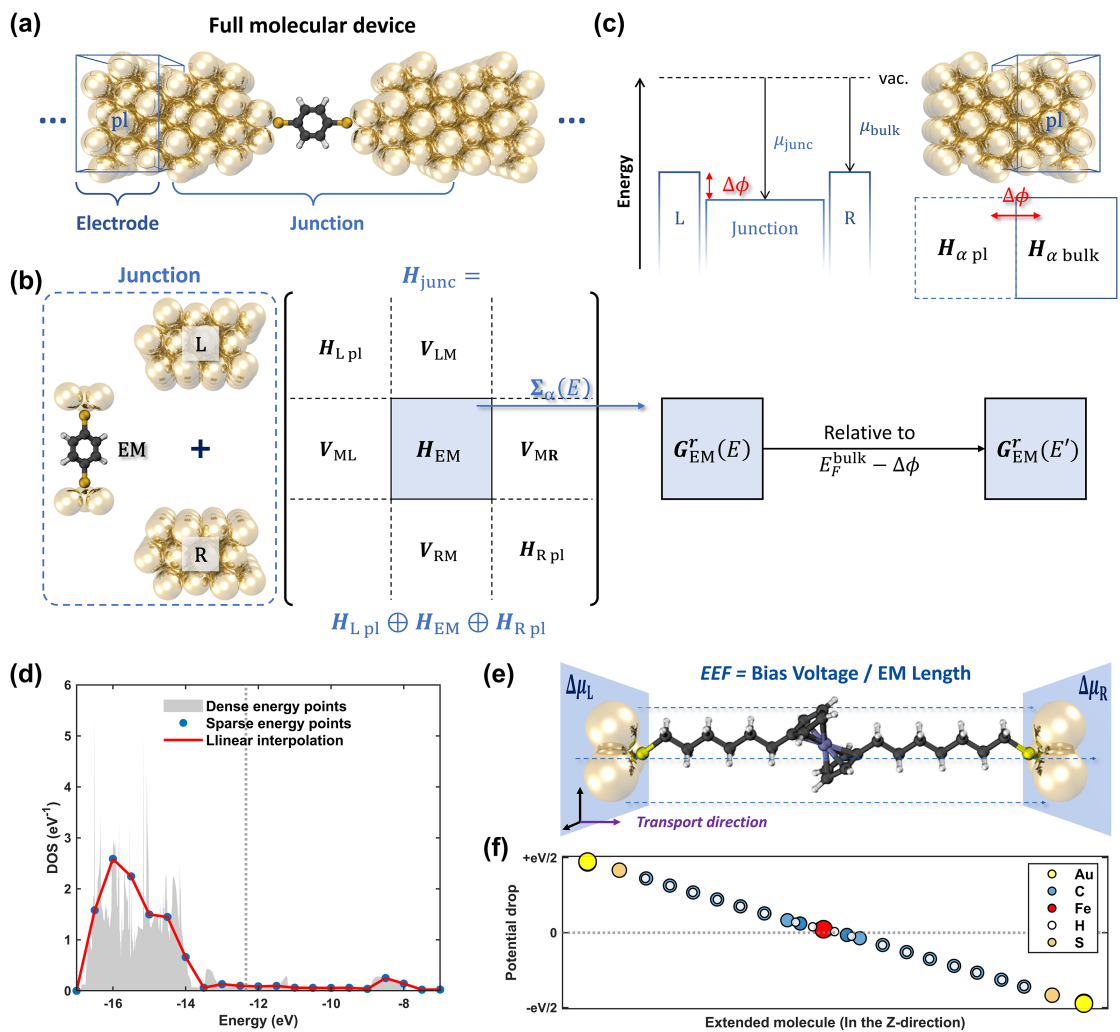

**Figure 4.** (a) Schematic diagram of a full molecular device where the extended molecule is embedded in the nanogap between semi-infinite electrodes, marked as the Level 3 (L3) scheme's two-step focus on the bulk-phase electrodes and the junction region. The solid box indicates a principal layer (pl) of the electrode. (b) Diagram illustrating the Hamiltonian partition for the junction system in Level 3. The extended molecule (EM) and one principal layer from each electrode from the junction, with the schematic presentation in Level 3 approach worked on the junction region, who are constructed with extended molecule and one principal layer connected at each side, are together computed for the Hamiltonian, and then partitioned into subregions. By taken the similar treatment in Level 2, the Green's function in the extended molecule size for transport calculation are obtained. the Level 3 (L3) approach, illustrating a full molecular junction where the extended molecule is connected to semi-infinite electrode principal layers on both sides. In practical calculations, the electrode and junction regions are computed separately, with corresponding matrix operations. (c) Energy-level diagram (left panel) showcases the chemical potential mismatching in the junction region with the bulk-phase electrode, and the matching of principal layer Hamiltonians in junction region and bulk Hamiltonians via diagonal element adjustments ($\Delta\phi$) to account for potential deviations. (d) Comparison of density of states (DOS) obtained from densely sampled energy points (gray shading) versus DOS linearly interpolated from surface Green's function (SGF) at sparse energy points (0.5 eV intervals, red line). (e) Schematic representation of the transport model under an applied bias voltage ($V$), simulating chemical potential changes in the left ($\Delta\mu_L$) and right ($\Delta\mu_R$) electrodes using a uniform extended electric field (EEF). The field strength is set to match the bias voltage across the length of the extended molecule, using a ferrocenyl-based system as an example. (f) Calculated potential drop along the transport direction in the extended molecule under applied bias, based on the diagonal elements of the Hamiltonian before and after the introduction of the EEF.

## 4.4. The QDHC Framework

Our computational framework, structured across three hierarchical levels, strikes a balance between accuracy and computational efficiency, enabling the analysis of systems with diverse characteristics at minimal cost. Each level progressively enhances the modeling of molecular transport systems by advancing computational methods, expanding system sizes, and refining the treatment of self-energy. This hierarchical design allows us to capture essential aspects of the charge transport, such as molecular quantum interference effects, interface interactions, energy level alignment, and current-voltage (I-V) characteristics under bias. Building upon this framework, we introduce the Question-Driven Hierarchical Computation (QDHC) strategy, which provides targeted guidance to the users in selecting the appropriate level based on their research objectives and specific transport properties of interest. Table 2 summarizes the three method and offers a comparative analysis of them within this framework, highlighting their respective strengths and suitable applications.

**Table 2.** Overview of Hierarchical Computational Levels in the QDHC Framework.

| Approaches | Transport Region | Approximation Techniques | Primary Insights |
|---|---|---|---|
| L1 | Isolated molecule | Multi-precision computation for molecular Hamiltonian; Self-energy under WBL approximation. | Transport by intrinsic molecular properties, like quantum interference, molecular conformational effects, etc. |
| L2 | Extended molecule with electrode clusters | Optimized electrode cluster size determined using a cutoff criterion; self-consistently interfacial coupling and self-energy based on a constant LDOS assumption for electrodes. | Molecule-electrode interface, like interfacial coupling, contact configuration, etc. |
| L3 | Extended molecule with electrode principal layers | Energy-dependent self-energy interpolated from iterated surface Green's functions at sparse energy points; linear potential polarization under bias. | Energy level alignment with respect to electrode's Fermi level; I-V characteristics (rectification, etc.) |

## 5. Results of Benchmark studies

To validate the versatility and effectiveness of our hierarchical framework, we performed benchmark studies on several representative molecular systems from the literature, each with increasing in complexity. These systems were carefully chosen to align with the three computational levels outlined above, encompassing a range of scenarios: from isolated molecules to extended system incorporating molecule-electrode interfaces, detailed electrode descriptions, and at last, the transport under bias. This diverse selection captures the wide variety of molecular junction configurations, demonstrating our framework's potential for broad applicability.

In the following sections, we detail the computational charge transport studies for each system, comparing our results with existing literature data. This comparative analysis provides a rapid and comprehensive assessment of key transport factors across different junction scenarios, showcasing the framework's ability to accurately model charge transport in various molecular junction systems.

5.1. L1: Molecule

At Level 1, we focus on the intrinsic properties of isolated molecules. The extended Hückel molecular orbital (EHMO) method can be used as a basic approximation, and it is particularly useful for initial explorations of molecular systems where detailed self-consistent calculations may not be necessary. For systems with significant intramolecular interactions or strong coupling effects, it is crucial to account for charge redistribution by employing self-consistent charge calculations for a more accurate Hamiltonian. In cases involving intricate electronic structures or quantum interference effects, even higher levels of computational precision and parameterization are required. To illustrate the application of our Level 1 framework, we present a typical example: Fano resonance in a para-carbazole anion junction. Two additional cases with varying computational precision demands are discussed in the Supplementary Information (Figure S10, S11).

Fano resonance, a distinctive quantum interference effect arising from interactions between molecular electronic states. In their 2022 study, Hong *et al.*[82] experimentally observed Fano resonance in a para-carbazole anion junction using electrochemical-gated STM-BJ. This scenario exemplifies the application of L1 scheme, where the electronic structure calculations were performed by the semi-empirical GFN1-xTB method that incorporates charge self-consistency, essential for accurate description of charge redistribution especially para-carbazole anion after deprotonation. Figure 5a illustrates the molecular structure before and after deprotonation, highlighting the localized negative charge at the carbazole center, which underscores the importance of precise treatment of charge rearrangement.

Using our Level 1 framework under the WBL approximation, with a coupling strength of 0.2 eV at the sulfur atoms, we successfully reproduced the characteristic antisymmetric peak in the transmission spectrum of the para-carbazole anion, see Figure 5b. Analysis of the eigen electronic structures reveals that the Fano resonance arise from interference between the localized HOMO, associated with the negative charge at the molecular center, and the delocalized LUMO. We further investigated how the Fano resonance responds to changes in the electrostatic environment by adjusting the position of a counterion near the charge center, thereby modifying electrostatic screening effects. As shown in Figure 5c, the Fano resonance can be finely tuned across specific energy ranges in the transmission spectrum by varying the distance between the counterion ($K^+$) and the nitrogen atom in the molecule.

This example demonstrates a typical case where the charge transport properties of the molecular junction are governed by the molecule's intrinsic electronic structure. By applying Level 1 scheme, we successfully reproduced results that align with both experimental observations and high level DFT+NEGF calculations, with a computation cost around 1~2 minutes using a common PC with a 6-core i7-10650H CPU. While referenced computations using the commercial program QuantumATK with GGA-PBE functional and mixed basis sets can only be performed by high-performance computer and are significantly more time-consuming.

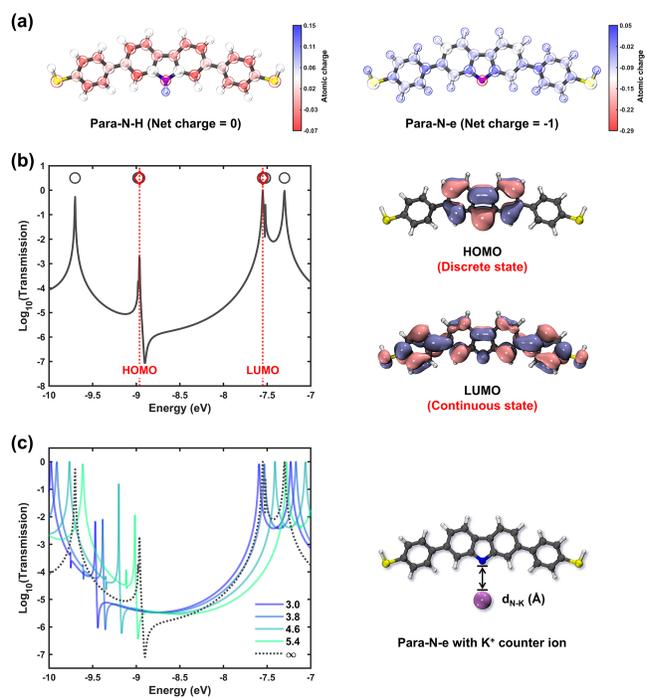

**Figure 5.** (a) Molecular representation of the para-carbazole molecule (Para-N-H) and its deprotonated form (Para-N-e) after the addition of a strong base, with atomic charge distributions visualized before and after deprotonation. The localization of negative charge on the carbazole center highlights the necessity of considering charge self-consistency in the computation. (b) Transmission spectrum of the Para-N-e calculated using EHMO with charge self-consistency, along with eigenstate plots for the marked resonance peaks. (c) Transmission spectra showing the regulating effect of varying the distance between the $K^+$ counterion and the nitrogen atom in the molecule.

## 5.2. L2: Extended molecule

Molecule-Electrode Interface

At level 2, we involve the effect of molecule-electrode interactions to account for the effect of conformation of molecules at the interface in transport properties. Venkataraman *et al.*[83] demonstrated in their 2009 study the conductance switching in the 4,4'-bipyridine junction through mechanical elongation and compression using an STM-BJ setup due to the variation in molecule-electrode interface contact.

To model this effect, we constructed extended bipyridine molecules linked to gold adatoms at varying tilt angles, using $Au_{15}$ clusters as electrodes according to Table 1 (Figure 6a). This computational model was optimized to represent the interfacial configuration proposed for pyridine-terminated molecules. Our model successfully captures the interaction between the molecule and the electrode. The resulting transmission spectra (Figure 6b) illustrate how interfacial coupling (evidenced by resonance peak broadening) and conductance at the Fermi level (approximated by the HOMO energy of the $Au_{15}$ cluster) differ between straight (90°) and tilted (50°) geometries during elongation process. Despite the rather rough approximation of the Fermi level position, it successfully reproduces the impact of molecule-electrode interface on conductance. These results demonstrate the effectiveness of our approach in accurately modeling interfacial effects and reinforce the reliability of our Level 2 scheme.

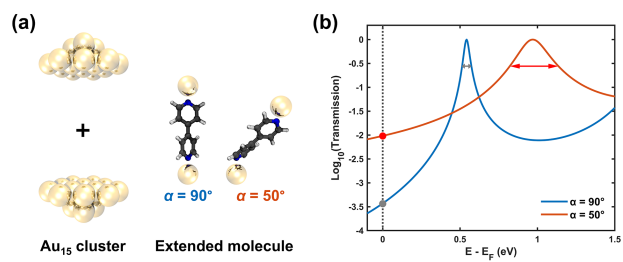

**Figure 6.** (a) Computational model for bipyridine junctions at Level 2, with the 15-atom Au cluster optimized for the adatom interfacial configuration, as determined in the previous section. (b) Comparative transmission spectra for the bipyridine molecule in contact with Au adatoms at different tilt angles α, corresponding to geometries reported in the literature[83], with the Fermi energy set to the HOMO energy of $Au_{15}$ cluster.

### 5.3. L3: Junction

Conducting Channel

Level 3 is essential when Fermi level alignment is required to determine the conducting channels. In their 2012 study, Venkataraman *et al.*[84] demonstrated how conductance and thermopower measurements, using a reformed STM-BJ setup, could be used in identifying dominant charge transport channels by analyzing Seebeck coefficients.

We reproduced their results by our Level 3 framework with two molecules: 4,4'-diaminostilbene (denoted as Amine) and 4,4'-bipyridine (denoted as Pyridine). They were constructed into molecular junctions (Figure 7a) by placing the electrode principal layer on both sides of the extended molecules (gold adatoms included), with geometries referenced from the literature.

As shown in Figure 7b, the transmission spectrum reveals that the Pyridine-terminated junction exhibits a resonance peak below the Fermi energy, while the Amine-terminated junction shows a resonance peak above. The proximity of these resonance peaks to the Fermi level indicates the dominant conducting channels. Eigen transport channel analysis at the resonance energies, as shown by iso-surfaces in Figure 7a, confirms that the Pyridine-terminated junction conducts via the LUMO, while the Amine-terminated junction conducts via the HOMO. It is important to point out that a further improvement on the quantitative energy alignment estimation, methods of higher precision or self-energy corrections[75–77] are necessary to account for many-body interactions, including precise exchange-correlation and image charge screening from the metal substrate.[75,85]

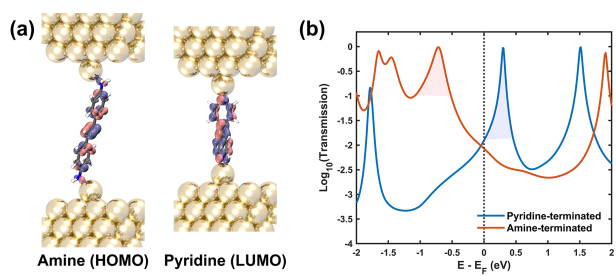

**Figure 7.** (a) Computational model of two junctions systems for Level 3, building extended molecules as referenced from the literature[83,84], linked to gold adatoms and connected to $3 \times 2\sqrt{3}$ Au(111) principal layers on each side. The iso-surfaces overlaid on the molecular structures represent the dominant molecular orbitals responsible for charge transport (HOMO for amine, LUMO for pyridine). (b) Comparative transmission spectra for pyridine and amine-terminated junctions, computed using the L3 framework. The shaded regions

highlight the resonance peaks corresponding to the dominant charge transport channels for each system.

Rectifier

We now extend the analysis to biased junction, to investigate non-equilibrium behaviors such as rectification. In their 2015 study, Yuan *et al.*[86] demonstrated a molecular rectifier where rectification can be adjusted by manipulating the position of the active center within the molecule.

To explore this phenomenon, we simulated two $SC_mFcC_nS$ junctions (Figure 8a) that differ in the position of the ferrocenyl (Fc) group within the nanogap, while keeping the total molecular length constant at $m + n = 13$. We then applied a uniform external electric field (EEF) along the transport direction during DFT calculation to generate the Hamiltonian. The resulting linear potential drop across the molecule acts as a good proxy for applying a finite bias voltage on a molecular device.[87] The field strength was determined by dividing the bias voltage by the length of extended molecule region. Its effect on the molecular electronic structure was illustrated in Figure 4d. Notably, under the same bias, the $SC_{10}FcC_3S$ molecule, which has greater structural asymmetry, shows a more pronounced energy level shift at the active ferrocenyl center. This shift causes resonance peaks to enter the bias window under one polarity more than the opposite polarity, leading to rectification. Correspondingly, transmission results in Figure 8 b and c indicate that the $SC_{10}FcC_3S$ junction, with stronger and reversed spatial asymmetry, undergoes a significant and opposite shift in its transmission spectrum compared with that of $SC_6FcC_7S$. This difference results in $SC_{10}FcC_3S$ displaying a larger and opposite rectification ratio. These findings are consistent with results from experimental study[86,88], the standard DFT+NEGF calculation[89], as well as single-state modeling under Landauer formula [33,88].

Compared with the typically high computational demands of investigating non-equilibrium transport behaviors using DFT+NEGF, our framework significantly reduces the computational load. This efficiency enables the analysis of transport behaviors of molecular junctions at each voltage point within minutes on a standard personal computer.

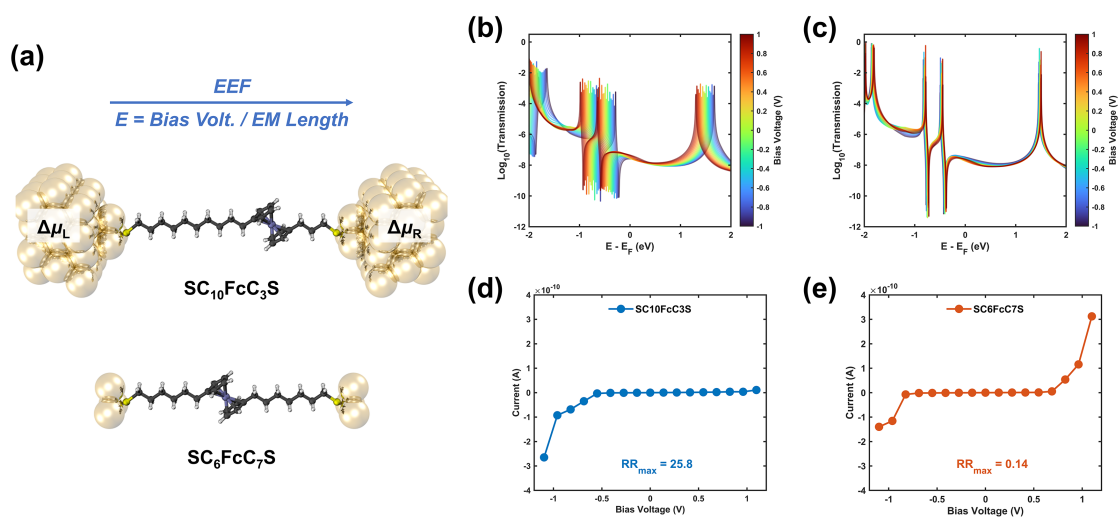

Figure 8. (a) Molecular junction models for Level 3 (biased scenario) computation, showing two ferrocenyl-based molecular junctions with the ferrocenyl (Fc) groups positioned differently within the nanogap, as referenced from the literature[86]. The sulfur atoms in the molecules are linked to Au trimers, forming the extended molecules, which are further connected to $3 \times 2\sqrt{3}$ Au(111) principal layers on each side. The biased condition in the device is implemented by shifting the chemical potentials of the left and right electrodes,

$\Delta\mu_L = eV/2$ and $\Delta\mu_R = -eV/2$, respectively. The resulting potential drop inside is approximated by applying a uniform external electric field (EEF) along the transport direction during the DFT calculation. The field strength is set by dividing the bias voltage by the length of the extended molecule region. The remaining computational procedures were carried out as previously described. Transmission spectra of (b) SC$_{10}$FcC$_3$S and (c) SC$_6$FcC$_7$S junctions under different bias voltages. Current-voltage (I-V) characteristics for (d) SC$_{10}$FcC$_3$S and (e) SC$_6$FcC$_7$S junctions, obtained by integrating the transmission functions over the respective voltage windows.

## 6. Conclusions

This work presents a three-tiered hierarchical approximation framework designed to model charge transport in molecular junctions across three critical aspects: system size, ab initio methods, and self-energy treatments. By scaling from isolated molecules to extended systems, and advancing from simple constant self-energy terms to more complex, energy-dependent treatments, this framework enables more efficient simulations of molecular junctions while maintaining a balance between computational cost and accuracy. Benchmark comparisons with established results validate the robustness of these approximations and demonstrate their utility for exploring charge transport in molecular devices.

The incorporation of the Question-Driven Hierarchical Computation (QDHC) strategy enhances the framework's versatility, particularly for analyzing larger systems such as supramolecular assemblies, biomacromolecules, and dynamic processes. Moreover, this framework establishes a foundation for future developments, including potential integration with emerging technologies such as deep-learning-assisted Hamiltonians and DFT+$\Sigma$ corrections. These advancements promise to further expand the framework's capabilities and applications in simulating charge transport across increasingly sophisticated molecular systems.

ASSOCIATED CONTENT

**Supporting Information**. The following files are available free of charge.

Generality of the localized coupling pattern for different interface configurations and proposed electrode cluster size cut-off criteria (Figure S1-S6). Computational time comparison for Level 2 scheme with varying electrode cluster sizes (Table S1, Figure S7). Electronic properties characterization of gold electrodes (Figure S8, S9). Two more benchmarking cases for Level 1 scheme (Figure S10, Figure S11). (PDF)

AUTHOR INFORMATION

**Corresponding Author**

* jixuan_0808@tju.edu.cn. * xi.yu@tju.edu.cn.

**Notes**

The authors declare no competing financial interest.

ACKNOWLEDGMENT

This project is supported by This work is supported by ISF-NSFC Joint Scientific Research Program (22361142833), Open Project of the State Key Laboratory of Supramolecular Structure and Materials (SKLSSM2024035), Open Project of the Key Laboratory of Resource Chemistry, Ministry of Education (2024-002003) and 'the Fundamental Research Funds for the Central Universities'.

ABBREVIATIONS

NEGF, Non-Equilibrium Green's Function; DFT, Density functional theory; WBL, Wide-band limit; QDHC, Question-driven hierarchical computation; SGF, Surface Green's function; EEF, External electric field.

# Table of Contents

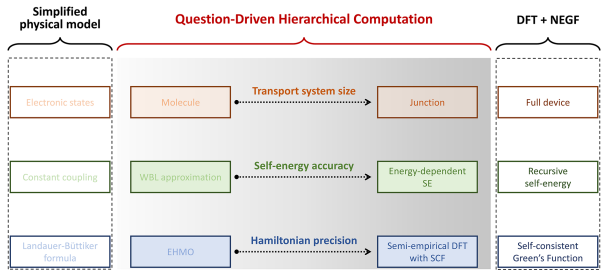